\documentclass[showpacs,twocolumn,aps]{revtex4}
\usepackage{epsf}
\begin{document}
\title{Exchange Driven Growth}
\author{E.~Ben-Naim}
\email{ebn@lanl.gov} \affiliation{Theoretical Division and Center
for Nonlinear Studies, Los Alamos National Laboratory,  Los
Alamos, New Mexico, 87545}
\author{P.~L.~Krapivsky}
\email{paulk@bu.edu} \affiliation{Center for Polymer Studies, and
Department of Physics, Boston University, Boston, MA, 02215}
\begin{abstract}
  We study a class of growth processes in which clusters evolve via
  exchange of particles. We show that depending on the rate of
  exchange there are three possibilities: I) Growth: Clusters grow
  indefinitely; II) Gelation: All mass is transformed into an infinite
  gel in a finite time; and III) Instant Gelation.  In regimes I and
  II, the cluster size distribution attains a self-similar form. The
  large size tail of the scaling distribution is
  $\Phi(x)\sim\exp(-x^{2-\nu})$, where $\nu$ is a homogeneity degree
  of the rate of exchange. At the borderline case $\nu=2$, the
  distribution exhibits a generic algebraic tail, $\Phi(x)\sim
  x^{-5}$. In regime III, the gel nucleates immediately and consumes
  the entire system. For finite systems, the gelation time vanishes
  logarithmically, $T\sim [\ln N]^{-(\nu-2)}$, in the large system
  size limit $N\to\infty$. The theory is applied to coarsening in the
  infinite range Ising-Kawasaki model and in electrostatically driven
  granular layers.

\end{abstract}
 \maketitle

\section{Introduction}

A multitude of growth phenomena in physical processes are driven by
exchange of particles between clusters.  Examples include droplet
growth via evaporation and re-condensation \cite{meakin}, island
growth in deposition processes \cite{az}, and phase ordering
\cite{ls,ajb,sm}. Exchange processes have been also used to model
social and economical systems including segregation of heterogeneous
populations \cite{ts}, the distribution of wealth in a society
\cite{ikr}, and growth of urban populations \cite{lr,klz}.

In exchange processes, clusters are composed of `atoms'
(monomers). Monomers detach from one cluster and re-attach to another
cluster. We shall consider the detachment controlled limit where the
time scale for transport between clusters is much faster then the time
scale for detachment.  Exchange processes incorporate both reversible
and irreversible features. Clusters may grow or shrink, yet when a
monomer attaches to another cluster, its respective cluster
disappears. This irreversible step provides the mechanism for cluster
growth.  Therefore, exchange-driven processes are fundamentally
different from irreversible growth processes, particularly aggregation
\cite{mvs,sc,skf}.

Such mass transfer processes are governed by an exchange kernel
$K(i,j)$ that represents the rate of transfer of monomers from a
cluster of size $i$ to a cluster of size $j$.  Generally, the rate of
monomer exchange between two clusters depends on their sizes.
Moreover, we consider the case where there is {\em no} preferable
direction for exchanges, i.e., symmetric exchange kernels,
$K(i,j)=K(j,i)$. This is unlike migration processes where the exchange
is preferential (``big gets bigger'' or ``rich gets
richer''). Migration underlies certain physical processes (e.g.,
coarsening with conserved order parameter \cite{ls,ajb}) as well as
social and economical processes \cite{ikr,lr,klz}.

We investigate homogeneous exchange kernels,
$K(ai,aj)=a^{2\lambda}K(i,j)$.  In particular, we consider the product
kernel $K(i,j)=(ij)^{\lambda}$ and its generalization $K(i,j)=i^\nu
j^\mu+i^\mu j^\nu$ with $\nu+\mu=2\lambda$ and $\nu\geq\mu$.  We
obtain a complete description of the problem in the asymptotic scaling
regime.  The overall range of possible behaviors and the emergence of
self-similar size distributions are as in aggregation and migration
processes. However, there are quantitative and qualitative
differences.  Unlike aggregation, the gelation transition is complete,
and unlike migration, the size distributions are extended rather than
compact.

The behavior falls into three categories. I) Growth: When $\nu<2$ and
$\lambda<3/2$, clusters grow indefinitely. The typical cluster size
grows algebraically with time, $k\sim t^{1/(3-2\lambda)}$, and the
cluster size distribution is given by a self-similar distribution with
a stretched exponential tail. II) Gelation: When $\nu<2$ and
$\lambda>3/2$, the entire mass in the system is transformed into an
infinite gel in a finite time. The cluster mass diverges algebraically
near the gelation point, $k\sim (t_c-t)^{1/(3-2\lambda)}$ and a
scaling behavior similar to the one underlying the growth phase is
found. In the borderline case $\nu=2$ the scaling function has an
algebraic tail with a universal exponent $\Phi(x)\sim x^{-5}$. Scaling
breaks down in the special point $\nu=\mu=2$ where the distribution is
log-normal. III) Instant gelation: When $\nu>2$, the gelation time
vanishes logarithmically with the system size, $t_c\sim [\ln
N]^{-(\nu-2)}$. In particular, for an infinite system, gelation is
instantaneous.

This paper is organized as follows. In the next section, we define the
exchange process. The governing equations are analyzed using scaling
techniques and exact solutions for the moments. We first analyze the
product kernel (Sec.~III) and then, the generalized kernel
(Sec.~IV). The gelation time in finite systems is investigated in
Sec.~V using heuristic arguments and numerical simulations.
Applications to coarsening in the Ising model with infinite range
Kawasaki dynamics and in electrostatically driven granular layers are
briefly discussed in Sec.~VI, and conclusions are given in Sec.~VII.

\section{Exchange Processes}

We consider the following elementary exchange process.  The system
consists of an ensemble of clusters and clusters evolve via transfer
of a single monomer from one cluster to another.  Symbolically,
\begin{equation}
\label{rule} (i,j) \buildrel K(i,j)\over \longrightarrow (i\pm
1,j\mp 1)
\end{equation}
with $i$ and $j$ the number of particles in each cluster and
$K(i,j)$ the exchange kernel. In an exchange event, a cluster is
equally likely to gain or to lose a particle. Since the exchange
process is unbiased, the matrix of transition rates is symmetric:
$K(i,j)=K(j,i)$.

Let $A_k(t)$ be the density of clusters containing $k$ monomers at
time $t$. It evolves according to the following rate equation
\begin{equation}
\label{rate-eq} \frac{dA_k}{dt}=\sum_{i,j}A_iA_j K(i,j)
\left[\delta_{k,i+1}+\delta_{k,i-1}-2\delta_{k,i}\right].
\end{equation}
This equation assumes perfect mixing, or equivalently, absence of
spatial correlations. We restrict our attention to monodisperse
initial conditions, $A_k(0)=\delta_{k,1}$.  The exchange process
has a single conservation law.  As reflected by the evolution
equations, the total mass is conserved, $M_1=1$ with $M_a=\sum_n
k^a A_k(t)$ the moments of the size distribution.  It is natural
to consider homogeneous kernels, $K(ai,aj)=a^{2\lambda}K(i,j)$
with $2\lambda$ the homogeneity degree and we present results for
the product kernel $K(i,j)=(ij)^{\lambda}$ and the generalized
homogeneous kernel $K(i,j)=i^\nu j^\mu+i^\mu j^\nu$ with
$\mu+\nu=2\lambda$.

\section{The product kernel}

For the product kernel, $K(i,j)=(ij)^\lambda$, the rate equations
(\ref{rate-eq}) read
\begin{eqnarray*}
\frac{dA_k}{dt}=M_\lambda \left[(k+1)^\lambda
A_{k+1}+(k-1)^\lambda A_{k-1}-2k^\lambda A_{k}\right]
\end{eqnarray*}
with the boundary condition $A_0\equiv 0$.  These evolution
equations demonstrate the diffusive character of the exchange
process. Absorbing the factor $M_\lambda$ into the time variable
\begin{equation}
\label{tau-I} \tau=\int_0^t dt'\,M_\lambda(t')
\end{equation}
we recast the governing equations into
\begin{equation}
\label{rate-eq-I} \frac{dA_k}{d\tau}= (k+1)^\lambda
A_{k+1}+(k-1)^\lambda A_{k-1}-2k^\lambda A_{k}\,.
\end{equation}
Alternatively, one can study integer moments of the size
distribution. The total density obeys $\frac{d}{d\tau}M_0=-A_1$,
the total mass is conserved $\frac{d}{d\tau}M_1=0$, and higher
integer moments satisfy the following hierarchy of equations
\begin{equation}
\label{mom-eq-I} \frac{d}{d\tau}M_n=2\sum_{l=1}^{[n/2]}{n\choose
2l}M_{n-2l+\lambda}.
\end{equation}
Only for integer values of the homogeneity index is this hierarchy
closed. We employ different approaches for different $\lambda$'s.  For
$\lambda<2$, we perform a scaling analysis of the rate equations and
for $\lambda\ge 2$, we analyze the moment equations.  This general
analysis is augmented by exact solutions for the integer values
$\lambda=0$, $1$, and $2$.

\subsection{Scaling ($\lambda<2$)}

When $\lambda<2$, dimensional analysis of Eq.~(\ref{rate-eq-I}) shows 
that the typical cluster size grows as
\begin{equation}
\label{scale-tau} k\sim \tau^\alpha,\qquad{\rm with}\qquad
\alpha=\frac{1}{2-\lambda}.
\end{equation}
Using $\frac{d\tau}{dt}=M_\lambda\sim \tau^{\alpha(\lambda-1)}$,
the growth of the typical scale is expressed in terms of the
physical time
\begin{equation}
\label{scale-t} k\sim \cases{t^{\beta}&$\lambda<3/2$,\cr \exp({\rm
const.}\times t)&$\lambda=3/2$,\cr
(t_c-t)^{\beta}&$3/2<\lambda<2$.}
\end{equation}
The dynamical exponent is $\beta=(3-2\lambda)^{-1}$.  As long as
$\lambda<3/2$, clusters grow indefinitely and the characteristic size
grows algebraically with time. For $\lambda>3/2$, a gelation
transition occurs, i.e., the system develops a giant cluster in a
finite time $t_c$.

We seek a scaling solution of the rate equations
\begin{equation}
\label{scaling-def} A_k(\tau)\simeq
\tau^{-2\alpha}\Phi\left(k\,\tau^{-\alpha}\right).
\end{equation}
Mass conservation dictates the normalization $J_1=1$ where $J_a=\int
dx\,x^a\,\Phi(x)$ is the $a^{\rm th}$ moment of the scaling
distribution. Technically, the scaling function describes the behavior
in the limits $k\to\infty$, $\tau\to\infty$ with the variable
$x=k\tau^{-\alpha}$ fixed.  Thus, we consider the continuum limit of
the rate equation \hbox{$\frac{\partial}{\partial
\tau}A(k,\tau)=\frac{\partial^2}{\partial k^2}\left[k^\lambda
A(k,\tau)\right]$}.  The scaling function satisfies the second order
linear differential equation
\begin{equation}
\label{scaling-eq-I} (2-\lambda)\,\frac{d^2}{dx^2}
\left[x^{\lambda}\Phi(x)\right]+x\frac{d}{dx}\Phi(x)+2\Phi(x)=0.
\end{equation}
Multiplying this equation by $x$, employing the identities
$x^2\Phi'+2x\Phi=(x^2\Phi)'$, $x\Psi''=(x\Psi)''-2\Psi'$, and
integrating once yields
\hbox{$(2-\lambda)\left[(x^{\lambda+1}\Phi)'-2x^\lambda
\Phi\right]+x^2\Phi(x)=0$}. Integrating a second time gives the
scaling function:
\begin{equation}
\label{scaling-sol-I}
\Phi(x)=C\,x^{1-\lambda}\exp\left[-\frac{x^{2-\lambda}}{(2-\lambda)^2}
\right]
\end{equation}
with
$C=(2-\lambda)^{-2/(2-\lambda)}/\Gamma\left(\frac{1}{2-\lambda}\right)$
found from the condition $J_1=1$.  The nature of the scaling function
differs from that found for migration where $K(l,m)=0$ for $l<m$
\cite{lr}: Exchange is characterized by extended distributions, while
migration is characterized by compact distributions.

There are two physically relevant cases for which the rate equations
can be solved exactly. When the exchange kernel is independent of the
cluster size ($\lambda=0$), the rate equation is
$\frac{d}{d\tau}A_k=A_{k+1}+A_{k-1}-2A_{k}$ and the cluster size
distribution is \cite{ikr}
\begin{equation}
\label{lambda=0}
A_k=e^{-2\tau}\left[I_{k-1}(2\tau)-I_{k+1}(2\tau)\right],
\end{equation}
where $I_n$ is the modified Bessel function of order $n$ \cite{bo}. In
agreement with the general scaling analysis, the typical scale grows
diffusively, $k\sim \tau^{1/2}$, and the scaling function is given by
$\Phi(x)=(4\pi)^{-1/2}x\,\exp(-x^2/4)$.

For the pure product kernel ($\lambda=1$), the rate equations read
$\frac{d}{d t}A_k=(k+1)A_{k+1}+(k-1)A_{k-1}-2kA_{k}$ (in this case
$t=\tau$).  Substituting the mass-conserving ansatz
$A_k=(1-u)^2u^{k-1}$ reduces the infinite set of rate equations
into a single ordinary differential equation $\frac{d}{dt}u=(1-u)^2$
subject the initial condition $u(0)=0$.  The size distribution in
this case is
\begin{equation}
\label{Ak1} A_k=\frac{t^{k-1}}{(1+t)^{k+1}}\,.
\end{equation}
The typical cluster size grows ballistically, $k\sim t$, and the
scaling function is purely exponential, $\Phi(x)=e^{-x}$, again in
agreement with the above scaling results.

When $3/2<\lambda<2$, an infinite cluster is formed at some finite
time $t_c$, termed the gelation time. The gelation time {\it
depends} on the initial condition and its determination requires
the full time dependent behavior.  Even without knowing the
gelation time exactly, one can describe the behavior in the pregel
stage since the size distribution still admits the scaling form
(\ref{scaling-sol-I}). Thus, for all $\lambda<2$ we have
\begin{equation}
\label{Pregel} A_k(\tau)\simeq
C\,k^{1-\lambda}\tau^{-\frac{3-\lambda}{2-\lambda}}
\exp\left[-\frac{k^{2-\lambda}\, \tau^{-1}}{(2-\lambda)^2} \right]
\end{equation}
implying that $A_k\to 0$ in the limit $\tau\to \infty$ ($t\to
t_c$). In other words, the gelation is complete, $A_k(t)=0$, for
$t\geq t_c$. This behavior differs from aggregation processes where at
the gelation time, all mass in the system is contained in finite size
clusters \cite{vE,van,FL}.

Complete gelation can be alternatively shown as follows. Let us assume
that the cluster size distribution approaches a constant $A_k\to
A_k^*>0$ as $\tau\to\infty$. From Eqs.~(\ref{rate-eq-I}), the
quantities $B_k\equiv k^\lambda A_k^*$ satisfy the discrete Laplace
equation $B_{k+1}+B_{k-1}-2B_{k}=0$ for $k>1$ and $B_2=2B_1$. Solving
recursively yields $B_k=kB_1=kA^*_1$ or
$A^*_k=k^{1-\lambda}A_1^*$. Mass conservation, $\sum_k kA_k^*=1$, 
implies $A^*_1=0$, and thence $A_k^*=0$ for all $k$, i.e., complete
gelation.

\subsection{Multiscaling ($\lambda=2$)}

In this special case, the moment equations (\ref{mom-eq-I}) are
closed for $n\ge 2$. For example
\begin{eqnarray}
\frac{dM_2}{d\tau}&=&2M_2,\nonumber\\
\frac{dM_3}{d\tau}&=&6M_3,\\
\frac{dM_4}{d\tau}&=&12M_4+2M_2.\nonumber
\end{eqnarray}
The solutions to these equations are combinations of exponentials:
$M_2=e^{2\tau}$, $M_3=e^{6\tau}$,
$M_4=\frac{6}{5}e^{12\tau}-\frac{1}{5}e^{2\tau}$, etc. The
physical time $t=\frac{1}{2}[1-e^{-2\tau}]$ is found from
$t=\int_0^\tau ds M_2^{-1}(s)$ so
\begin{eqnarray}
M_2&=&(1-2t)^{-1},\nonumber\\
M_3&=&(1-2t)^{-3},\\
M_4&=&\frac{6}{5}(1-2t)^{-6}-\frac{1}{5}(1-2t)^{-1}.\nonumber
\end{eqnarray}
Therefore, the gelation time is $t_c=1/2$.  Asymptotically, the
first term in (\ref{mom-eq-I}) dominates:
\hbox{$\frac{d}{d\tau}M_n\simeq n(n-1)M_n$} implying $M_n\sim
\exp[n(n-1)\tau]$ for $n>1$. Close to the gelation time ($t\to
t_c$), the moments diverge according to
\begin{equation}
\label{mom-gen-I} M_n(t)\sim (t_c-t)^{-n(n-1)/2}.
\end{equation}
In this special case, moments exhibit multiscaling asymptotic
behavior, namely, the normalized moments $M^{1/n}_n/M_1$ diverge.

To determine the asymptotic form of the size distribution we treat $k$
as a continuous variable. For $\lambda=2$, Eq.~(\ref{rate-eq-I})
becomes \hbox{$\frac{\partial}{\partial \tau}
A_k=\frac{\partial^2}{\partial k^2} [k^2A_k]$}. This equation is
equidimensional in $k$ \cite{bo} thereby suggesting use of the
variable $\xi=\ln k$ instead of $k$.  Making the transformation from
$A_k(t)$ to $A(\xi,\tau)$ defined via $A_k dk=A(\xi)d\xi$, we recast
above equation for $A_k(t)$ into the following constant coefficients
diffusion-convection equation
\begin{equation}
\left(\frac{\partial}{\partial \tau}-\frac{\partial}{\partial
\xi}\right) A(\xi,\tau)= \frac{\partial^2}{\partial
\xi^2}A(\xi,\tau).
\end{equation}
With the initial conditions $A(\xi,0)=\delta(\xi)$, the solution
reads $A(\xi,\tau)=(4\pi\tau)^{-1/2}\exp[-(\xi+\tau)^2/(4\tau)]$.
The original distribution $A_k=k^{-1}A(\xi)$ is log-normal
\begin{equation}
\label{log-normal} A_k(\tau)\simeq
(4\pi\tau)^{-1/2}e^{-\tau/4}k^{-3/2} \exp\left[-\frac{(\ln
k)^2}{4\tau}\right].
\end{equation}
Again, the distribution vanishes at the transition point, i.e.,
the gelation transition is complete. Moreover, the mass
distribution is algebraic, $A_k(t)\sim M_0(t)\,k^{-3/2}$ for
sufficiently small masses, $k\ll \sqrt{\ln\frac{1}{1-2t}}$.  The
total density vanishes quite slowly near the transition point
\begin{equation}
M_0(t)\sim (1-2t)^{1/8}\left(\ln \frac{1}{1-2t}\right)^{-1/2}.
\end{equation}
We note that the density follows a different law than the one
characterizing higher than first moments (\ref{mom-gen-I}).

The size distribution does not follow a scaling behavior
asymptotically and the log-normal distribution is responsible for the
multiscaling behavior (\ref{mom-gen-I}) of the moments. This differs
from aggregation processes where the moments diverge as $M_n(t)\propto
(t_c-t)^{-\alpha_n}$ \cite{vE} with the exponent $\alpha_n$ {\em
linear} in $n$.

\subsection{Instant Gelation ($\lambda>2$)}

Gelation is now instantaneous and complete, that is $A_k(t)=0$ for all
$k$ when $t>0$.  To prove this assertion we assume the opposite and
arrive at a contradiction.  Our analysis follows van Dongen's approach
developed in the context of aggregation processes \cite{van}.

The moments $M_n$ with integer $n\geq 2$ evolve according to
(\ref{mom-eq-I}).  The first term in the summation yields a lower
bound for their growth rate
\begin{equation}
\label{Mn} \frac{dM_n}{d\tau}\geq n(n-1)M_{n-2+\lambda}\geq
n(n-1)(M_n)^{1+\Lambda}
\end{equation}
with $\Lambda=\frac{\lambda-2}{n-1}$. The second inequality
follows from the Jensen's inequality as shown below.  Consider the
auxiliary functions ${\cal M}_n$, evolving according to
\begin{eqnarray}
\label{calM} \frac{d {\cal M}_n}{d\tau}=n(n-1) ({\cal
M}_n)^{1+\Lambda}\,.
\end{eqnarray}
Solving this equation subject to the initial condition ${\cal
M}_n(0)=1$ yields ${\cal M}_n=[1-n(\lambda-2)\tau]^{-1/\Lambda}$.
Therefore, ${\cal M}_n\to\infty$ as
$\tau\to\tau_n=n^{-1}(\lambda-2)^{-1}$. Since $M_n\geq {\cal M}_n$,
the moment $M_n$ diverges at least at $\tau_n$. The series of times
$\tau_n$ set an upper bound for the gelation time $\tau_c$ since all
moments should be finite for $\tau<\tau_c$.  As $\tau_n\to 0$ when
$n\to\infty$, we conclude that $\tau_c=0$ and thence, the gelation
time vanishes $t_c=0$.

The inequality $M_{n-2+\lambda}\geq (M_n)^{1+\Lambda}$ with 
$\Lambda=\frac{\lambda-2}{n-1}$ is derived as follows.  Let the
parameters $p_j\geq 0$ satisfy $\sum_j p_j=1$ and let $\Phi(x)$ be a
convex function. A convex function satisfies the Jensen inequality
\begin{eqnarray}
\label{jensen} \sum_{j=1}^\infty p_j \Phi(x_j)\geq
\Phi\bigg(\sum_{j=1}^\infty p_j x_j\bigg).
\end{eqnarray}
First, we substitute the coefficients $p_j=jA_j$ (from mass
conservation $\sum_j jA_j=1$) and the convex function
$\Phi(x)=x^{1+\Lambda}$ ($\Lambda>0$ for $\lambda>2$) into the Jensen
inequality. Then, choosing $x_j=j^{n-1}$ and using $\sum p_jx_j=\sum
j^nA_j=M_n$ and $\sum p_j\Phi(x_j)=M_{n-2+\lambda}$ yields the above
inequality.

\section{Generalized kernels}

The rates $K(i,j)$ underlying exchange processes are typically
homogeneous functions of $i$ and $j$ (at least for large $i$ and
$j$). We restrict ourselves to such kernels. Apart from the
homogeneity degree $2\lambda$, homogeneous kernels are characterized
by an additional exponent $\nu$ defined through the asymptotic
$K(1,j)\sim j^\nu$ as $j\gg 1$.  For $i\ll j$ the exchange kernel
scales as $K(i,j)=i^{2\lambda} K(1,j/i)\sim i^\mu j^\nu$ with
$2\lambda=\nu+\mu$. Therefore, we consider a specific generalization
of the product kernel that exhibits these homogeneity properties
\begin{equation}
\label{kernel-II}
K(i,j)=i^\nu j^\mu+i^\mu j^\nu\,.
\end{equation}
More precisely, the asymptotics $K(i,j)\sim i^\mu j^\nu$ occurs for
$i\ll j$ if $\nu\geq \mu$; since the kernel is symmetric, we can
assume that $\nu\geq \mu$ without loss of generality. We expect that
the homogeneity indices govern the overall qualitative behavior
(growth, gelation, instant gelation), while the precise form of the
kernel controls quantitative characteristics such as the size
distribution.

For this exchange kernel, the rate equations (\ref{rate-eq}) become
\begin{eqnarray*}
\frac{d A_k}{dt}&=& M_\mu\left[(k+1)^{\nu} A_{k+1}+(k-1)^{\nu}
A_{k-1}-2k^{\nu} A_{k}\right]
\nonumber\\
&+& M_\nu\left[(k+1)^{\mu} A_{k+1}+(k-1)^{\mu} A_{k-1}-2k^{\mu}
A_{k}\right].
\end{eqnarray*}
The following generalization of the modified time variable
\begin{equation}
\label{tau-II} \tau=\int_0^t dt'\sqrt{M_\nu(t')M_\mu(t')}
\end{equation}
handles the two indices symmetrically. In terms of this time, the
evolution equations are
\begin{eqnarray*}
\frac{dA_k}{d\tau}&= &R\left[(k+1)^{\nu}
A_{k+1}+(k-1)^{\nu}A_{k-1}-2k^{\nu} A_{k}\right]
\\
&+& R^{-1}\left[(k+1)^{\mu} A_{k+1}+(k-1)^{\mu}
A_{k-1}-2k^{\mu}A_{k}\right], \nonumber
\end{eqnarray*}
with $R=\sqrt{M_\mu/M_\nu}$.  Of course, the dynamics conserve mass:
$\frac{d}{d\tau}M_1=0$. Higher integer moments evolve according to
\begin{equation}
\label{mom-eq-II} \frac{dM_n}{d\tau}=2\sum_{l=1}^{[n/2]}{n\choose
2l} \left[RM_{n-2l+\nu}+R^{-1}M_{n-2l+\mu}\right].
\end{equation}

When $\nu<2$, the scaling analysis follows closely the product kernel
case. The overall growth laws (\ref{scale-tau}) and (\ref{scale-t})
remain unchanged and the homogeneity degree $\lambda$ characterizes
the scaling behavior. However, we shall see that the individual
indices $\nu$ and $\mu$ play an important role since they dictate the
range for which this law holds.

We seek a scaling solution of the form (\ref{scaling-def}). The
scaling function $\Phi(x)$ satisfies
\begin{equation}
\label{scaling-eq-II}
\frac{d^2}{dx^2}\Big[\big(Ux^{\nu}+Vx^{\mu}\big)\Phi(x)\Big]+
x\,\frac{d}{dx}\Phi(x)+2\Phi(x)=0
\end{equation}
with the constants $U=\alpha^{-1}A$ and $V=\alpha^{-1}A^{-1}$
determined from the ratio $A=\sqrt{J_\mu/J_\nu}$. The scaling function
reads
\begin{equation}
\label{scaling-sol-II}
\Phi(x)=C\,\frac{x}{Ux^{\nu}+Vx^{\mu}}\,\exp\left[\!-\!\int_0^x\!dy\,
\frac{y}{Uy^{\nu}+Vy^{\mu}}\right].
\end{equation}

The scaling solution involves three parameters $U$, $V$, and $C$.
Substituting $U=\alpha^{-1}A$ and $V=\alpha^{-1}A^{-1}$ into the
equality $A=\sqrt{J_\mu/J_\nu}$ yields a closed equation for the
parameter $A$. Once $A$ is determined, the parameters $U$ and $V$
follow, and finally, the amplitude $C$ is found from the normalization
$J_1=1$.

We illustrate this procedure for the special case $(\nu,\mu)=(1,0)$,
i.e., for the pure sum kernel $K(i,j)=i+j$. In this case, the integral
on the right-hand side of Eq.~(\ref{scaling-sol-II}) is readily
computed. Using $U=\frac{3A}{2}$ and $V=\frac{3}{2A}$ we arrive at
\begin{equation}
\label{scal-special}
\Phi(x)=C\,x\left(1+A^2x\right)^{a-1}\exp\left[-aA^2 x\right]
\end{equation}
with $a=\frac{2}{3}A^{-3}$. We now substitute this solution into the
right-hand side of the equality $A=\sqrt{J_0/J_1}$ and transform it 
into the transcendental equation 
\hbox{$(\frac{e}{a})^a\,\Gamma(a,a)+a^{-1}=1$} involving the incomplete gamma
function (see Appendix A). The amplitude is then explicitly evaluated
to give $C=aA^6$. Numerically, we find $a\cong 2.82649$, $A\cong
0.428397$, and $C\cong 0.0174713$.  Interestingly, there is a
nontrivial algebraic correction to the leading exponential behavior,
$\Phi(x)\sim x^a\exp(-aAx)$ for large $x$.

On the boundary $\nu=2$ separating regime III from the two other
regimes, the solution of Eq.~(\ref{scaling-sol-II}) significantly
simplifies.  We find $A=1/[2(2-\mu)]$, $U=1/4$, $V=1/(2-\mu)^2$, and
consequently, the scaling function
\begin{equation}
\label{scal-boundary} \Phi(x)=C\,x^{1-\mu}
\left[1+\frac{x^{2-\mu}}{4(2-\mu)^2}\right]^{-1-\frac{4}{2-\mu}}.
\end{equation}
The constant $C=2\left[2(2-\mu)\right]^{-1-\frac{2}{2-\mu}}\,
[B(\frac{1}{2-\mu},\frac{3}{2-\mu})]^{-1}$ is expressed in terms
of the beta function.  Remarkably, the scaling function
(\ref{scal-boundary}) exhibits a universal large-$x$ asymptotic
behavior
\begin{equation}
\label{tail} \Phi(x)\sim x^{-5}\,.
\end{equation}
In other words, the size distribution is algebraic, $A_k(\tau)\sim
k^{-5}\,\tau^{3\alpha}$ where $\alpha=(2-\lambda)^{-1}=2/(2-\mu)$.
With this algebraic divergence, sufficiently small moments are
characterized by ordinary scaling behavior while higher moments exhibit
multiscaling behavior:
\begin{equation}
\label{mom-lead-II}
M_n\sim \cases {\tau^{\alpha(n-1)} &$n<4$,\cr
\tau^{\alpha n(n-1)/4}             &$n>4$.}
\end{equation}
This behavior follows from the leading term in the moment equation
(\ref{mom-eq-II}), viz.  $\frac{d}{d\tau}M_n=n(n-1)M_nR$.  With
$R=\sqrt{M_\mu/M_2}\simeq A\tau^{-1}$ and $A=\alpha/4$, this equation
becomes $\frac{d}{d\tau}M_n=\frac{\alpha n(n-1)}{4\tau}\,M_n$, leading
to the multiscaling behavior (\ref{mom-lead-II}).

The determination of $A$ in the general situation requires numerical
evaluation, yet the form and nature of the size distribution is
clear. For example, the minimal (maximal) index governs the
distribution of small (large) clusters. Indeed, from
Eq.~(\ref{scaling-sol-II}), the extremal behaviors are
\begin{equation}
\label{scaling-lim-II} \Phi(x)\sim \cases {x^{1-\mu}&$x\ll 1$,\cr
\exp[-x^{2-\nu}]&$x\gg 1$.}
\end{equation}
Apart from the point $(\nu,\mu)=(2,2)$, the scaling solution holds for
all $\nu\leq 2$. As in the product kernel case, growth occurs when
$\lambda<3/2$ and gelation occurs when $3/2<\lambda<2$.

For $\nu>2$, the scaling solution (\ref{scaling-sol-II}) predicts
$\Phi\sim x^{1-\nu}$. Such behavior is inconsistent since the moment
$J_\mu$ diverges and instead, instantaneous gelation occurs. The
moments $M_n$ with $n>1$ satisfy Eq.~(\ref{mom-eq-II}) and the first
term in the summation yields a lower bound for the moment growth
$\frac{d}{d\tau}M_n\geq R n(n-1)M_{n-2+\nu}$.  Keeping only this term
and absorbing the factor $R$ into the time variable, the previous
proof applies. Thus, gelation is instantaneous.

\begin{figure}
\centerline{\epsfxsize=7.6cm\epsfbox{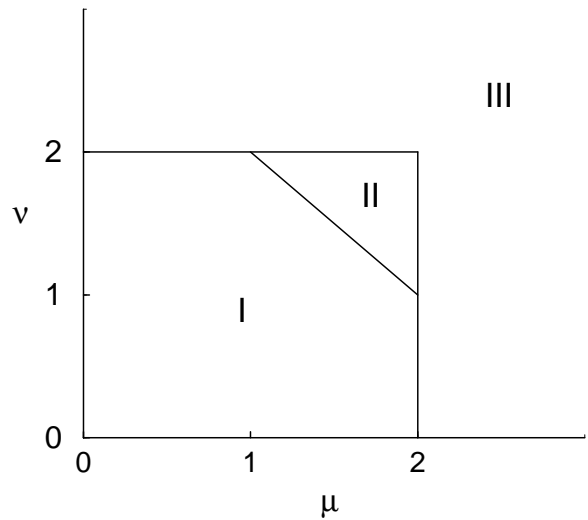}} \caption{The three
types of behaviors: Scaling (I), Ordinary gelation (II), and
Instant gelation  (III).} \label{types}
\end{figure}

To summarize, there are 3 types of behaviors, determined by the
homogeneity degrees $\mu$ and $\nu$ (Fig.~1):
\begin{enumerate}

\item[I] {\em Growth}.  The cluster size grows indefinitely, and the
size distribution obeys scaling.

\item[II] {\em Gelation}. The cluster size diverges in a finite time
and the size distribution follows a scaling solution near the gelation
time. Gelation is complete.

\item[III] {\em Instant Gelation}. The cluster size distribution
vanishes for all $t>0$.

\end{enumerate}

The cluster size distribution exhibits a scaling behavior in regimes I
and II.  Scaling behavior underlies the system everywhere except for
regime III and the point $(2,2)$. In the bulk of regimes I and II the
size distribution is a stretched exponential, while in the boundary
with region III, the cluster size distribution has an algebraic tail.
Last, at the point $(\mu,\nu)=(2,2)$ scaling breaks down and the
distribution is log-normal.

\section{The gelation time}

Instantaneous gelation is certainly counter intuitive: A finite time
singularity that occurs at time $t=+0$\,! Instantaneous gelation was
investigated exclusively in the context of aggregation
\cite{FL,dom,hez,spouge,van,jeon,mg}.  For infinite systems, it is
impossible to quantify the difference between two instant gelling.
Finite systems, on the other hand, naturally quantify how fast a
system gels.

Consider a system consisting initially of $N$ monomers. In a finite
time $t_N$, all mass in the system condenses into a single `runaway'
cluster. How does the average gelation time $T_N=\langle t_N\rangle$
depend on $N$? When growth or ordinary gelation occurs, the answer
follows from our previous analysis.  In the scaling regime, the growth
law (\ref{scale-t}) indicates that the condensation time grows
algebraically with the system size, $T_N\sim N^{1/\beta}$.  In the
case of ordinary gelation, the average gelation time saturates at an
$N$-independent value: $T_N\to t_c$. The interesting case is instant
gelation where the gelation time vanishes in the thermodynamic limit,
$T_N\to 0$ as $N\to \infty$.

For simplicity, we discuss the product kernel. The vanishing
gelation time is ultimately related to the short time behavior.
Early on, loss terms in the rate equation (\ref{rate-eq-I}) are
negligible and to leading order $\frac{d}{dt}A_j\cong (j+1)^\lambda 
A_{j+1}$, where we tacitly assumed $\tau\equiv t$. For the initial
condition $A_j(0)=\delta_{j,1}$, the leading order behavior of the
density is
\begin{equation}
A_{j+1}\cong (j!)^{\lambda-1}\,t^j.
\end{equation}
In a finite system consisting initially of $N$ monomers, a $j$-mer
first appears at time $t_j\approx (j!)^{-(\lambda-1)/j} N^{-1/j}$,
estimated from $NA_j(t_j)=1$. For example, the first dimer and trimer
appear at times $t_2=N^{-1}$ and $t_3=2^{-(\lambda-1)/2} N^{-1/2}$,
respectively. By definition, the times increase monotonically,
$t_{j+1}>t_j$, yet the above estimates increase monotonically only for
sufficiently small $j<j_*$. From $t_{j_*}=t_{j_*+1}$, we obtain the
extremum $j_*=(\lambda-1)^{-1}\ln N$ using the Stirling formula. The
corresponding time $T_*\equiv t_{j_*}$ is
\begin{equation}
\label{T*} T_*\sim \left(\frac{\lambda-1}{\ln
N}\right)^{\lambda-1}\,.
\end{equation}

For later times, $t\gg T_*$, the rate equations should be modified to
account for the finiteness of the system (see e.g.  \cite{aal,tn})
since significant statistical fluctuations are induced by large
runaway clusters that take over (eventually only one such cluster
remains). The critical size of such clusters is $j_*\sim
(\lambda-1)^{-1}\ln N$.  As a complete analytical solution seems out
of reach, we proceed heuristically by focusing on the leading cluster
that eventually grows to be the gel. Since it exchanges monomers back
and forth with other clusters its growth mechanism is diffusive.  For
an ordinary diffusive process, $\frac{d}{dt}\,\langle k\rangle=0$,
while \hbox{$\frac{d}{dt}\,\langle k^2\rangle=D$}. In our case,
$D=k^\lambda$ with the typical size $k^2\equiv \langle k^2 \rangle$.
Therefore, the typical size of the runaway cluster grows according to
\begin{equation}
\label{mass} \frac{dk}{dt}=k^{\lambda-1}.
\end{equation}
Integrating this rate equation from the critical size $k=j_*$ to the
system size $k=N$ gives the gelation time
\begin{equation}
\label{TF} T_N=T_*+\frac{1}{\lambda-2}
\left[\frac{1}{j_*^{\lambda-2}}-\frac{1}{N^{\lambda-2}}\right].
\end{equation}
Since $j_*=(\lambda-1)^{-1}\ln N$, the duration of the latter growth
phase is much larger than that of the nucleation phase, $T_N\gg
T_*$. Therefore, the gelation time vanishes logarithmically
\begin{equation}
\label{TN} T_N\sim (\ln N)^{-(\lambda-2)}\,,
\end{equation}
in the thermodynamic limit. A straightforward extension of the above
argument to the generalized exchange kernel (\ref{kernel-II}) gives
$T_N\sim (\ln N)^{-(\nu-2)}$.

\begin{figure}
\centerline{\epsfxsize=7.6cm\epsfbox{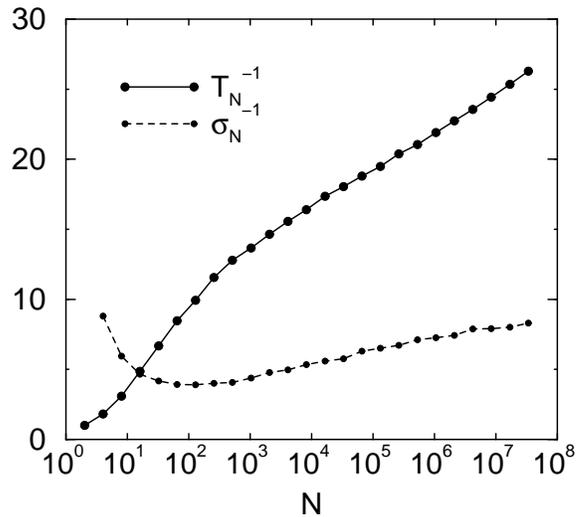}} 
\caption{The system size dependence of the gelation time. Shown are
the average gelation time $T_N$ and the normalized variance $\sigma_N$
versus the system size. The Monte Carlo simulation results correspond
to an average over $10^3$ independent realizations of the exchange
process with $\lambda=3$.}
\end{figure}

Therefore, in a finite system it may be difficult to distinguish
instantaneous gelation from the ordinary one.  We verified the
logarithmic law (\ref{TN}) numerically for $\lambda=3$ (Fig.~2).
Probing fluctuations in the gelation time, we observe that the
normalized variance $\sigma_N^2=\langle t_N^2\rangle/\langle
t_N\rangle^2-1$ vanishes logarithmically in the thermodynamic limit
(Fig.~2).  The distribution of normalized gelation times becomes
trivial $P(t_N/T_N)\to \delta(z-1)$ implying that the gelation time is
a self-averaging quantity.

We also examined the gelation time in two other growth processes,
namely aggregation \cite{mvs,sc,skf} and addition \cite{bk,l}. The
above heuristic picture yields a similar logarithmic law albeit with a
different exponent \cite{mg}. Self-averaging is observed numerically
as well, and we conclude that the behavior found for exchange processes
is generic.

\section{Applications to Coarsening}

In exchange processes, a monomer detaches from a cluster and
subsequently re-attaches to another cluster. This elementary mechanism
underlies a number of growth and coarsening processes.  We apply our
general theory to two coarsening processes.

\subsection{Infinite range Ising-Kawasaki model}

The zero temperature Ising model with infinite range Kawasaki dynamics
evolves via exchanges of spins of opposite signs belonging to domain
walls \cite{kk,tk,ab}. This model is equivalent to the Shelling's
segregation model \cite{ts}. In the limit of a vanishing volume
fraction of one of the two phases, the domains of the minority phase
are isolated and the process is essentially an exchange process with a
product kernel $K(i,j)=(ij)^\lambda$. The dependence of the number of
exchange candidates (i.e., spins in domain walls that can lower their
energy by hoping to a different cluster) on the cluster size dictates
the homogeneity degree.  For spherical clusters, only perimeter spins
may exchange. Since the island size and the surface size grow with the
radius according to $k\sim R^d$ and $\sigma \sim R^{d-1}\sim
k^{(d-1)/d}$, respectively, one has $\lambda=(d-1)/d$ and $\beta=
1/(3-2\lambda)=d/(d+2)$. The dynamical exponent (defined through
$R\sim t^z$) is therefore $z=1/(d+2)$. If the islands are polygons, a
distinct possibility on a lattice, then only corner spins are active
so $\lambda=0$, and consequently $\beta=1/3$ and $z=1/(3d)$. Both
estimates agree in one dimension, consistent with the exact solution
$z=1/3$ \cite{bk1}.

\subsection{Coarsening of thin granular layers}

In electrostatically driven granular layers, clusters nucleate around
large grains \cite{ia}. Then, as charged grains oscillate back and
forth between the two bounding plates due to the oscillating electric
field, they may scatter of the plate or collide with other
particle. Consequently, individual grains may transfer from one
cluster to another. The rate of hoping into and out of a cluster is
proportional to its area.  Therefore, the homogeneity degree is unity,
$\lambda=1$, implying $\beta=1$ and a dynamical exponent of
$z=1/d$. In two dimensions, this prediction is consistent with the
experimental observations $z=1/2$ \cite{ia}.  The corresponding size
distribution $\Phi(x)=e^{-x}$ provides a reasonable approximation for
the experimental size distribution, obtained from a relatively small
number of clusters (see Fig.~3).  The normalized variance, defined via
$\sigma^2=J_2/J_1^2-1$, is experimentally found to be $\sigma=0.80\pm
0.05$ compared with the theoretical value $\sigma=1$.  Comparing
"spatial" exchange processes where grains are exchanged only between
neighboring clusters with further experimental data may help elucidate
the relevance of spatial correlations.

\begin{figure}
\centerline{\epsfxsize=7.6cm\epsfbox{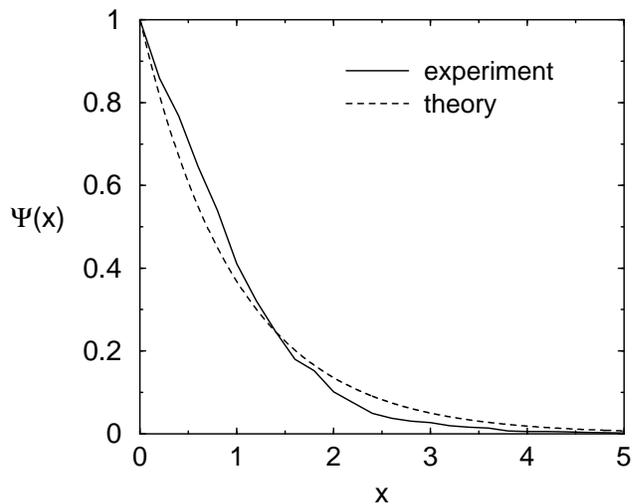}} 
\caption{The cumulative cluster area distribution, defined via
\hbox{$\Psi(x)=\int_x^\infty dy \Phi(y)$} versus the normalized area
$x$.  The experimental distribution represents roughly $10^3$ clusters
obtained from $20$ different snapshots during a single realization of
the coarsening process in which the total number of clusters evolved
from $200$ to $10$. Once the average area is set to unity, the
cumulative area distributions at different snapshots are the same.
The theoretical distribution $\Psi(x)=\Phi(x)=\exp(-x)$ follows from
Eq.~(\ref{Ak1}).}
\end{figure}

\section{Conclusions}

We have shown that kinetics of exchange processes are classified by
the homogeneity indices of the governing rates. There are three
possible regimes including indefinite growth, gelation in a finite
time, and instant gelation. Scaling behavior underlies the first two
regimes. The size distributions are generally extended, decaying
exponentially or algebraically for large sizes, in contrast with
migration processes.

We also studied the gelation time in finite systems and found that it
decays rather slowly, following an inverse logarithmic law. It would
be interesting to determine the full time dependent behavior of
moments of the size distribution in the instant gelation regime.  In
the realm of aggregation, instantaneous gelation is relevant in
astrophysics, e.g., to the coalescence of planetesimals to form
planets or of stars to form super-massive black holes (see \cite{mg}
and references therein). The physical relevance of instant gelation in
exchange processes has yet to be demonstrated.

Our description was on a mean-field level where all pairs of clusters
in the system are equally likely to interact. It will be interesting
to incorporate spatial fluctuations into this description. The nature
of the spatial fluctuations depends on the mechanism for transporting
monomers from one cluster to the other.  For diffusive transport, one
can incorporate effective fluxes into clusters, using the standard
techniques developed for reaction-diffusion processes.

\acknowledgements \noindent We thank Igor Aronson and Avner Peleg for
useful discussions.  We are grateful to Igor Aronson for providing the
experimental data.  This research was supported by DOE
(W-7405-ENG-36).

\appendix
\section{The case $(\nu,\mu)=(1,0)$}

Substituting Eq.~(\ref{scal-special}) into $A^2=J_0/J_1$ yields
\begin{eqnarray*}
1&=&\frac{\int_0^\infty dx\, x^2 \,(1+x)^{b-1}\,e^{-ax}}
{\int_0^\infty dx\, x\, (1+x)^{b-1}\,e^{-ax}}
\end{eqnarray*}
with $b=a$. Evaluation of the ratio of the integrals is performed
as follows
\begin{eqnarray*}
1&=&-\frac{d}{da}\ln
\left[\int_0^\infty dx\, x\, (1+x)^{b-1}\,e^{-ax}\right]\Big|_{a=b}\\
&=&-\frac{d}{da}\ln \left[-\frac{d}{da}
\left(\int_0^\infty dx\, (1+x)^{b-1}\,e^{-ax}\right)\right]\Big|_{a=b}\\
&=&-\frac{d}{da}
\ln\left[-\frac{d}{da}\left(e^aa^{-b}\,\Gamma(b,a)\right)\right]\Big|_{a=b}\\
&=&\frac{a^{-2}+e^aa^{-a-1}\,\Gamma(a,a)}{a^{-1}}.
\end{eqnarray*}

\end{document}